\documentclass{desyproc}

\begin{document}

\def\beq{\begin{equation}}
\def\eeq{\end{equation}}
\def\gtsim{\stackrel{>}{_\sim}}
\newcommand{\form}[1]{(\ref{#1})}
\newcommand{\gag}{g_{a\gamma\gamma}}

\title{Polarization mesurements of gamma ray bursts and axion like
particles}

\author{{\slshape Andr\' e Rubbia$^1$, Alexander Sakharov$^{1,2}$}\\[1ex]
$^1$Swiss Institute of Technology, ETH-Z\"urich, 8093 Z\"urich, Switzerland\\
$^2$TH Division, PH Department, CERN, 1211 Geneva 23, Switzerland }

\contribID{sakharov\_alexander \\ CERN-PH-TH/2008-187}

\desyproc{DESY-PROC-2008-02}
\acronym{Patras 2008} 
\doi  

\maketitle
\vspace*{-0.5cm}
\begin{abstract}
A polarized gamma ray emission spread over a sufficiently wide energy
band from a strongly magnetized astrophysical object like gamma ray bursts
(GRBs) offers an opportunity to test the hypothesis of axion like particles
(ALPs). Based on evidences of polarized gamma ray emission detected in
several gamma ray bursts we estimated the level of ALPs induced dichroism, which
could take place in the magnetized fireball environment of a GRB. This allows to
estimate the sensitivity of polarization measurements of GRBs to the ALP-photon
coupling. This sensitivity $\gag\le
2.2\cdot 10^{-11}\ {\rm GeV^{-1}}$ calculated for the ALP mass
$m_a=10^{-3}~{\rm eV}$ and MeV energy spread of gamma ray emission is
competitive with the sensitivity of CAST and becomes
even stronger for lower ALPs masses.
\end{abstract}

New very light spin-zero particles are predicted in many extentions of the
Standard Model (see this proceedings for the references). Typically, such
particles called axion like particles (ALPs) can
arrise as a result of a spontaneous brekdown of a continuose symmetry. A notable
example of such brekdown is  the Peccei-Quinn (PQ) mechanism~\cite{PQ}, which
remains perhaps the most natural solution to the CP problem in QCD. The
most important phenomenological 
property of ALPs is their two-photon vertex
interaction, which
allows for ALP to photon conversion in the presence of an external
electric and
magnetic fields~\cite{axion_gamma_vertex} through an interaction term
\beq
\label{gamma_vertex}
{\cal L}_{a\gamma}=-\frac{1}{4}\gag F_{\mu\nu}\tilde F^{\mu\nu}a=\gag{\bf
E\cdot B}a,
\eeq 
where $a$ is the ALP field, $F$ is the electromagnetic field strength tensor,
$\tilde F$ its dual, ${\bf E}$, ${\bf B}$ the electric and magnetic fields
respectively and  $\gag$ is the ALP-photon coupling strength. 

According to~\cite{mimmo} the ALP-photon mixing~\form{gamma_vertex} gives
rise to vacuum dichroism. 
This dichroism results in the rotation of the polarization plane of an
initially linearly polarized
monochromatic beam by angle given in~\cite{mimmo,raffelt}:
\beq
\label{rotation}
\epsilon=\frac{\gag^2B^2\omega^2}{m_a^2}
\sin^2\left(\frac{m_a^2L}{4\omega}\right)\sin 2\phi .
\eeq
It is valied for a uniform magnetic field ${\bf B}$ laing at a nonvanishing
angle $\phi$ with the wave vector ${\bf k}$ of photons with frequency $\omega$.
Here $m_a$ is the mass of ALP, $L$ is the length of the magnetized
region. Such rotation, for instance, in case of ALPs,   could be detected in a
laser experiment like PVLAS~\cite{pvlas1,pvlas2}. The validity of the
approximation~\form{rotation} is provided if
the oscillation wavenumber 
\beq
\label{osc_num}
\Delta_{\rm osc}^2=\left(\frac{{m_a^2}-\omega_{\rm pl}}
{2\omega}\right)^2+B^2\gag^2
\eeq  
is dominated by the axion mass term. In fact,~\form{osc_num} pertains to the
situation in which the beam propagates in a magnetized
plasma, which gives rise to an effective photon mass set by the plasma
frequency $\omega_{\rm pl}=\sqrt{4\pi\alpha n_e/m_e}\simeq 3.7\cdot
10^{-11}\sqrt{n_e/{\rm cm}^{-3}}$~eV, where $n_e$ is the electron density and
$m_e$ is the electron mass.

 The polarization of the prompt gamma ray emission has been measured in four
bright GRBs: GRB021206, GRB930131, GRB960924 and GRB041219a. The first
measurements made in~\cite{boggs}  with Ranaty High Energy Solar Spectrometer
Imager (RHESSI) satellite, found a linear polarization,
$\Pi =(80\pm 20)\%$, of the gamma rays from GRB021206 across the spectral window
0.15-2~MeV. The analysis techniques have been challenged in~\cite{critic} and
defended in~\cite{defend}. Subsequent analyses made in~\cite{psi} confirmed the
results of~\cite{boggs} but at the lower level of significance. Later,
in~\cite{batsepol} the BATSE instrument on board of the Compton Gamma Ray
Observatory (CGRO) has been used to measure, for two GRBs, the
angular distribution of gamma rays back-scattered by the rim of the Earth's
atmosphere: $35\%\le\Pi\le 100\%$ for GRB930131 and $50\%\le\Pi\le 100\%$ for
GRB960924. The analysing technique of~\cite{batsepol} is only sensitive to the
energy range 3-100~keV. Finally, the analysis~\cite{intpol} of GRB041219a
across the spectral window 100-350~keV has been performed using coincidence
events in the SPI (spectrometer on board of the INTEGRAL
satelite) and IBIS (the Imager on Board of the INTEGRAL
satelite). The polarization fraction of $\Pi =96^{+39}_{-40}\%$ was determined
 for this GRB. 
 
According to the 
Hillas~\cite{hillas} diagram showing size and magnetic field strengths of
different astrophysical
object the typical magnetic field in a GRB's engine can be estimated as
$B\simeq 10^9$~G over a region $L_{\rm GRB}\simeq
10^9$~cm. Moreover, the conservation of magnetic field energy at the rest wind
frame of fireball shell model of the GRB's engine~\cite{piran_pr} implies at any
radial distance $r$, in the fireball environment, $4\pi r_0^2B_0^2=4\pi r^2B^2$,
leading to the relation $B=B_0(r_0/r)$, where $B_0$ and $r_0$ are the magnetic
field strength and the size of the central part of the fireball. The
minimal time scale of variability of GRBs light curves is estimated
to be about 0.1 sec~\footnote{See, for example, the analysis in~\cite{wave}.}.
This implies that the typical extention of the GRB's engine is indeed
compatible with $L_{GRB}\approx 10^9$~cm. Typically the central
part of the fireball can be represented  by a neutron star of radius $r_0\approx
10^6$~cm with magnetic field of $B_0\approx 10^{12}$~G. Therefore the
strength of the magnetic field at the distance $r=L_{GRB}$ corresponds to 
$B\approx 10^9$~G, which is in a good agreement
with the values taken from~\cite{hillas}. 

According to~\form{rotation}, the relative misalignment between the polarization
planes of gamma radiation at two different energies $\omega_1$ and  $\omega_2$
induced by ALPs (see for details~~\cite{sr}) is given by
\beq
\label{relative}
\Delta\epsilon =\frac{L_{GRB}}{2\pi}\frac{\gag^2}{m_a^2}\Delta\omega B^2,
\eeq
where $\Delta\omega=|\omega_2-\omega_1|$. Therefore, one can observe that the
constraint arises from the fact that if the ALP dichroism induced rotation of
polarization plane~\form{relative} in the given magnetic field were to differ
by more then
$\pi /2$ over the energy range 0.2-1.3~MeV, as in the case of GRB021206, the
instantaneous polarization in the detector would fluctuate significantly for the
net time averaged polarization of the signal to be suppressed. This condition
can be transformed into the bound on the ALP-photon coupling as
\beq
\label{b1}
\gag\le\pi\frac{m_a}{B\sqrt{\Delta\omega L_{\rm GRB}}}\approx 2.2\cdot
10^{-8}\frac{m_a}{1\ {\rm eV}}\ ({\rm GeV})^{-1},
\eeq 
\begin{figure} [t]
\begin{center}
\epsfig{file=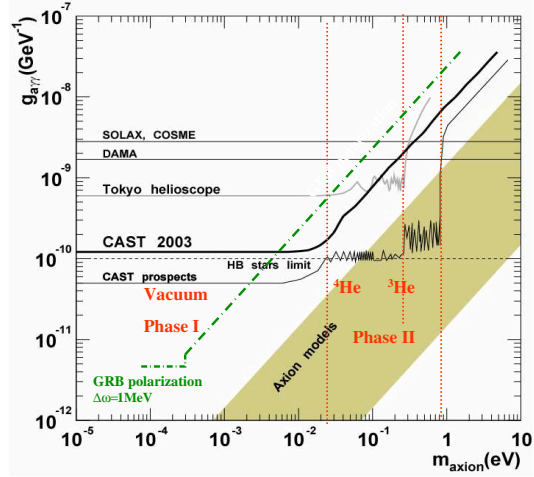,width=70mm,clip=}
\end{center}
\caption{\it The plot of the regions of $(m_a, \gag)$ space 
ruled out by various solar axion searches with
the bound of the present letter, estimated for the inner part of the energy
range 0.2-1.3~MeV applied for the polarization measurements of
GRB021206 (dashed
dotted line), superimposed.}
\label{castlimit}
\end{figure}
where the inner part of the spectral window 0.2-1.3~MeV ($\Delta\omega\approx
1{\rm MeV}$)
reported in polarization analysis of GRB021206 has been used. This constraint
is obtained under the assumption that the correlation length of the magnetic
field being initially defined by the typical size of the neutron star in the
core of a GRB's engine is getting streched out by the axpansion of the fireball
shell. So, at some moment of the expansion the correlation length becomes
adjusted
to the oscillation length. However, for the ALP's mass
\beq
\label{m_crit}
m_a\le m_{cr1}=\sqrt{\frac{2\pi\omega}{L_{GRB}}}\approx 3.5\cdot 10^{-4}\ {\rm
eV}
\eeq 
this condition does not hold anymore and the polarization planes
misalignment angle
should be calculated as
\beq
\label{miss2}
\Delta\epsilon =B^2\gag^2L_{\rm GRB}\left(\frac{L_{\rm
GRB}}{16}-\frac{\omega_1}{2\pi
 m_a^2}\right).
\eeq
  The expression~\form{miss2} 
holds to be positive down to the
mass (see for details~\cite{sr}): $m_{\rm cr2}= 4\sqrt{\frac{\omega_1}{2\pi
L_{GRB}}}\approx 8\cdot 10^{-5}\ {\rm
eV}$.
Requiring again that the misalignment angle~\form{miss2} does not exceed $\pi
/2$ in the axion mass range between $m_{\rm cr1}$ and $m_{\rm cr2}$ one
arrives to a bound, which can be well approximated by a constant~\footnote{The
electron number density
in a GRB's
environment can be estimated as $n_e\simeq 10^{10}\ {\rm
cm}^{-3}$~\cite{piran_pr}. Therefore the
expression~\form{osc_num} is still ALP mass dominated down to
$m_a\approx m_{\rm cr}$ for the energy of the gamma
radiation, $\omega\approx 1$~MeV, and constraints on $\gag$
calculated from~\form{b1} and~\form{lim_low_mass}. }
\beq
\label{lim_low_mass}
\gag\le \frac{2\sqrt{2\pi}}{BL_{\rm GRB}}\approx 5\cdot 10^{-12}\ ({\rm
GeV})^{-1}. 
\eeq

In Fig.~1. we show the bounds~\form{b1} and~\form{lim_low_mass}
superimposed on the recent results of CAST~\cite{cast} and other axion
helioscope experiments~\cite{rev_exp}. The limit obtained
becomes by factor
$\sqrt{1{\rm MeV}/\Delta\omega_{I,B}}$ weaker if we apply the width
$\Delta\omega_{I}\approx 250$~keV of the energy bands for GRB041219a 
 detected by INTEGRAL or $\Delta\omega_{B}\approx 100\ {\rm keV}$\- for
GRB930131 and
GRB960924 detected by BATSE. This implies that $\gag\le
4.4\cdot 10^{-11}\ {\rm GeV^{-1}}$ and $\gag\le
6.9\cdot 10^{-11}\ {\rm GeV^{-1}}$ for INTEGRAL and BATSE measurements
 respectively calculated for the axion mass
$m_a=10^{-3}~{\rm eV}$.  

An improovement of the current estimations could be achived in further
detection of gamma polarized signals from GRBs in the similar or higher energy 
ranges. For thir reasons the~POLAR \cite{polar} experiment as well as other
numerous efforts to develop instruments
with the sensitivity required for astrophysical polarimetry over 100~eV to
10~GeV band~\cite{pol_astro} become important probes for ALPs beyond Standard
Model physics. 
\\
\\
We thank K.~Zioutas, H.~Hofer and N.~Produit for useful discussions on CAST,
POLAR and related topics.

\begin{footnotesize}

\end{footnotesize}

\begin{thebibliography}{99}


\bibitem{PQ} R.D.~Peccei and H.~Quinn, Phys. Rev. Lett. {\bf 38} (1977) 1440

\bibitem{axion_gamma_vertex} D.~A.~Dicus, E.~W.~Kolb, V.~L.~Teplitz and
R.~V.~Wagoner,
  Phys.\ Rev.\  D {\bf 18} (1978) 1829; P.~Sikivie,
  Phys.\ Rev.\ Lett.\  {\bf 51} (1983) 1415
  [Erratum-ibid.\  {\bf 52} (1984) 695].

\bibitem{mimmo} L.~Maiani, R.~Petronzio and E.~Zavattini,
  Phys.\ Lett.\  B {\bf 175} (1986) 359.
  
\bibitem{raffelt} P.~Sikivie,
  Phys.\ Rev.\ Lett.\  {\bf 51} (1983) 1415
  [Erratum-ibid.\  {\bf 52} (1984) 695];
  G.~Raffelt and L.~Stodolsky,  Phys.\ Rev.\  {\bf 37} (1988) 1237.
 
\bibitem{pvlas1}
  E.~Zavattini {\it et al.}  [PVLAS Collaboration],
  Phys.\ Rev.\ Lett.\  {\bf 96} (2006) 110406
  [arXiv:hep-ex/0507107].
  
\bibitem{pvlas2} E.~Zavattini {\it et al.}  [PVLAS Collaboration],
  arXiv:0706.3419 [hep-ex].

\bibitem{boggs} W.~Coburn and S.E.~Boggs, Nature {\bf 423}, 415 (2003).
    

\bibitem{critic} R.~E.~Rutledge and D.~B.~Fox,
  Mon.\ Not.\ Roy.\ Astron.\ Soc.\  {\bf 350} (2004) 1272
  [arXiv:astro-ph/0310385].
  
\bibitem{defend}   S.~E.~Boggs and W.~Coburn,
  arXiv:astro-ph/0310515.
  
\bibitem{psi}   C.~Wigger, W.~Hajdas, K.~Arzner, M.~Gudel and A.~Zehnder,
  Astrophys.\ J.\  {\bf 613} (2004) 1088
  [arXiv:astro-ph/0405525].
  
\bibitem{batsepol}   D.~R.~Willis {\it et al.},
  arXiv:astro-ph/0505097.
  

\bibitem{intpol}   S.~McGlynn {\it et al.},
  arXiv:astro-ph/0702738.


  
\bibitem{boggs_lv} S.~E.~Boggs, C.~B.~Wunderer, K.~Hurley and W.~Coburn,
  Astrophys.\ J.\  {\bf 611} (2004) L77
  [arXiv:astro-ph/0310307].
  
\bibitem{hillas}   A.~M.~Hillas,
  Ann.\ Rev.\ Astron.\ Astrophys.\  {\bf 22} (1984) 425; L.~Anchordoqui,
T.~Paul, S.~Reucroft and J.~Swain, 
  Int.\ J.\ Mod.\ Phys.\  A {\bf 18} (2003) 2229
  [arXiv:hep-ph/0206072].
  
\bibitem{piran_pr} T.~Piran,
  Phys.\ Rept.\  {\bf 314} (1999) 575
  [arXiv:astro-ph/9810256].

\bibitem{wave}  
  J.~R.~Ellis, N.~E.~Mavromatos, D.~V.~Nanopoulos and A.~S.~Sakharov,
  Astron.\ Astrophys.\  {\bf 402} (2003) 409
  [arXiv:astro-ph/0210124].
  
\bibitem{sr}   A.~Rubbia and A.~S.~Sakharov,
  Astropart.\ Phys.\  {\bf 29} (2008) 20.
 
    
\bibitem{cast}
  S.~Andriamonje {\it et al.}  [CAST Collaboration],
  JCAP {\bf 0704} (2007) 010
  [arXiv:hep-ex/0702006]; K.~Zioutas {\it et al.}  [CAST Collaboration],
  Phys.\ Rev.\ Lett.\  {\bf 94} (2005) 121301
  [arXiv:hep-ex/0411033].

\bibitem{rev_exp} For review see: R.~Battesti {\it et al.},
  arXiv:0705.0615 [hep-ex]; 
D. Lazarus et al., Phys. Rev. Lett. {\bf 69} (1992) 2089; 
S. Moriyama et al., Phys. Lett. {\bf B434} (1998) 147; 
R. Bernabei et al., Phys. Lett. {\bf B515} (2001) 6; 
F.T. Avignone et al., Phys. Rev. Lett. {\bf 81} (1998) 5068;
R.J. Creswick et al., Phys. Lett. {\bf B427} (1998) 235.

\bibitem{polar} N.~Produit {\it et al.},
  Nucl.\ Instrum.\ Meth.\  A {\bf 550} (2005) 616
  [arXiv:astro-ph/0504605]; 
  

\bibitem{pol_astro} For reviews see: J.K.~Black, 3rd Sympothium on Large TPCs
for Low Energy Rare Event Detectors, Journal of Physics, Conference Series {\bf
65} (2007) 012005; A.~Curioni, Ph.D. Dissertation Thesis, Columbia
University (2004), unpublished; 
A.~Rubbia,
 J.\ Phys.\ Conf.\ Ser.\ {\bf 39}, 129 (2006)
 [arXiv:hep-ph/0510320].
\end{thebibliography}
\end{document}